\documentclass[twocolumn]{jpsj3}
\usepackage{txfonts}
\RequirePackage{graphicx,color}

\title{
Theory of Spin-Acoustic Resonance for Spin-3/2 Si Vacancy with $C_{3v}$ Site Symmetry
in Silicon Carbide
}

\author{Mikito Koga$^1$ and Masashige Matsumoto$^2$}

\inst{
$^1$Department of Physics, Faculty of Education, Shizuoka University, Shizuoka
422--8529, Japan \\
$^2$Department of Physics, Faculty of Science, Shizuoka University, Shizuoka
422--8529, Japan \\
}

\abst{
Motivated by the recent acoustically driven spin resonance studies applied to silicon vacancy
centers in silicon carbide, we theoretically investigate the spin--strain interaction characterized by
the defect spin-$3/2$ quadrupole components coupled to strain fields.
Considering the $C_{3v}$ symmetry of the vacancy site beyond the spherical approximation, we
clarify the effect of a deviation from the spherical symmetry on spin resonance transition rate,
which can be changed by rotating a static magnetic field.
The ratios of
spin--strain coupling parameters can be evaluated from the anisotropic field-direction
dependence of the transition rate using a standing or traveling surface acoustic wave.
We also discuss the effect of the propagation direction of the acoustic wave tilted from the
crystallographic mirror plane reflecting the $C_{3v}$ symmetry.
The results presented here reveal the quadrupole properties inherent in spin-3/2 states 
and will promote the realization of the acoustically driven strain control of spin.
}

\begin{document}

\maketitle

%%%%%%%%%%%%%%%%%%%%%%%%%%%%%%%%%%%%%%%%%%%%%%%%%%%%%%%%%%%%%%%%%%%%%%%%%%%%%%%%
%Macros
%%%%%%%%%%%%%%%%%%%%%%%%%%%%%%%%%%%%%%%%%%%%%%%%%%%%%%%%%%%%%%%%%%%%%%%%%%%%%%%%
\newcommand{\ds}{\displaystyle}

\renewcommand{\H}{\mathcal{H}}
\newcommand{\br}{{\mbox{\boldmath$r$}}}
\newcommand{\bR}{{\mbox{\boldmath$R$}}}
\newcommand{\bS}{{\mbox{\boldmath$S$}}}
\newcommand{\bk}{{\mbox{\boldmath$k$}}}
\newcommand{\bH}{{\mbox{\boldmath$H$}}}
\newcommand{\bh}{{\mbox{\boldmath$h$}}}
\newcommand{\bJ}{{\mbox{\boldmath$J$}}}
\newcommand{\bI}{{\mbox{\boldmath$I$}}}
\newcommand{\bPsi}{{\mbox{\boldmath$\Psi$}}}
\newcommand{\bpsi}{{\mbox{\boldmath$\psi$}}}
\newcommand{\bPhi}{{\mbox{\boldmath$\Phi$}}}
\newcommand{\bd}{{\mbox{\boldmath$d$}}}
\newcommand{\bG}{{\mbox{\boldmath$G$}}}
\newcommand{\bu}{{\mbox{\boldmath$u$}}}
\newcommand{\be}{{\mbox{\boldmath$e$}}}
\newcommand{\bE}{{\mbox{\boldmath$E$}}}
\newcommand{\bp}{{\mbox{\boldmath$p$}}}
\newcommand{\bB}{{\mbox{\boldmath$B$}}}
\newcommand{\om}{{\omega_n}}
\newcommand{\omm}{{\omega_{n'}}}
\newcommand{\omd}{{\omega^2_n}}
\newcommand{\omt}{{\tilde{\omega}_{n}}}
\newcommand{\ommt}{{\tilde{\omega}_{n'}}}
\newcommand{\btau}{{\hat{\tau}}}
\newcommand{\brho}{{\mbox{\boldmath$\rho$}}}
\newcommand{\bsigma}{{\mbox{\boldmath$\sigma$}}}
\newcommand{\bSigma}{{\mbox{\boldmath$\Sigma$}}}
\newcommand{\bt}{{\hat{t}}}
\newcommand{\bq}{{\hat{q}}}
\newcommand{\bLambda}{{\hat{\Lambda}}}
\newcommand{\bDelta}{{\hat{\Delta}}}
\newcommand{\bU}{{\hat{U}}}
\newcommand{\bskp}{{\mbox{\scriptsize\boldmath $k$}}}
\newcommand{\skp}{{\mbox{\scriptsize $k$}}}
\newcommand{\bsrp}{{\mbox{\scriptsize\boldmath $r$}}}
\newcommand{\bsRp}{{\mbox{\scriptsize\boldmath $R$}}}
\newcommand{\bsk}{\bskp}
\newcommand{\sk}{\skp}
\newcommand{\bsr}{\bsrp}
\newcommand{\bsR}{\bsRp}
\newcommand{\ri}{{\rm i}}
\newcommand{\re}{{\rm e}}
\newcommand{\rd}{{\rm d}}
\newcommand{\rM}{{\rm M}}
\newcommand{\rs}{{\rm s}}
\newcommand{\rt}{{\rm t}}
\newcommand{\Tc}{{$T_{\rm c}$}}
\renewcommand{\Pr}{{PrOs$_4$Sb$_{12}$}}
\newcommand{\La}{{LaOs$_4$Sb$_{12}$}}
\newcommand{\LaPr}{{(La$_{1-x}$Pr${_x}$)Os$_4$Sb$_{12}$}}
\newcommand{\PrLa}{{(Pr$_{1-x}$La${_x}$)Os$_4$Sb$_{12}$}}
\newcommand{\OsRu}{{Pr(Os$_{1-x}$Ru$_x$)$_4$Sb$_{12}$}}
\newcommand{\PrRu}{{PrRu$_4$Sb$_{12}$}}
%%%%%%%%%%%%%%%%%%%%%%%%%%%%%%%%%%%%%%%%%%%%%%%%%%%%%%%%%%%%%%%%%%%%%%%%%%%%%%%%%%%%%%%%%%%%%%%%%%%

%%%%%%%%%%%%%%%%%%%%%%%%%%%%%%%%%%%%%%%%%%%%%%%%%%%%%%%%%%%%%%%%%%%%%%%%%%%%%%%%%%%%%%%%%%%%%%%%%%%
\section{Introduction}
%%%%%%%%%%%%%%%%%%%%%%%%%%%%%%%%%%%%%%%%%%%%%%%%%%%%%%%%%%%%%%%%%%%%%%%%%%%%%%%%%%%%%%%%%%%%%%%%%%%
High spin ($S \ge 1)$ states can be coupled to strain fields driven by acoustic waves or mechanical
oscillators.
\cite{VanVleck40,Donoho64,Udvarhelyi18a}
The mechanical control of spin provides another possibility of quantum spin control at the nanoscale,
as extensively studied using optically and magnetically controlled spin systems such as the
nitrogen-vacancy (NV) centers in diamond.
\cite{Udvarhelyi18a,MacQuarrie13,Kepesidis13,Ovartchaiyapong14,MacQuarrie15,Barfuss15,Meesala16,Golter16,Degen17,Lee17,Chen18,Barfuss19,Barry20,Chen20,Norman21}
The silicon vacancy (V$_{\rm Si}$) centers in silicon carbide (SiC) are considered leading platforms for spin--strain coupling with high sensitivity to realize the acoustic control of quantum spin states, which can be considered an alternative to spin manipulation by microwaves.
\cite{Whiteley19a,Soltamov19,Whiteley19b,Singh22}
Unlike the NV spin-1 defect, the higher spin $S = 3/2$ of V$_{\rm Si}$ causes more 
complicated spin dynamics, and it is highly desirable to clarify the details of an acoustically driven
spin--strain interaction with the $C_{3v}$ site symmetry of V$_{\rm Si}$.
\cite{Mizuochi03,Isoya08,Simin16,Soykal17,Tarasenko18,Poshakinskiy19,Castelletto20,Sosnovsky21,Hernandez-Minguez21,Vasselon22,Dietz22}
\par

Recently, spin-acoustic resonance (SAR) measurements in 4H-SiC have shown that
phonon-driven quantum spin resonance transitions depend on the rotation of a static
magnetic field on the SiC surface.
\cite{Hernandez-Minguez20}
The SAR is driven by single and double quantum spin transitions between two different
$S = 3/2$ energy levels $m_S$ with $\Delta m_S = \pm 1$ and $\Delta m_S = \pm 2$.
Although the data analysis is based on a spherically symmetric spin--strain interaction model,
\cite{Poshakinskiy19,Breev21}
it is important that an observed nontrivial field-direction dependence should reflect the
symmetries of $S = 3/2$ quadrupole components coupled to strain fields driven by a surface
acoustic wave (SAW).
This is a key to our theory considering the $C_{3v}$ site symmetry of a defect spin beyond
the spherical approximation, which was previously proposed for phonon-driven resonance in the
NV centers.
\cite{Koga20,Koga22}
\par

In this paper, we present a useful representation of the spin--strain interaction for the V$_{\rm Si}$
spin $S = 3 /2$ under the rotation of a magnetic field and derive a fundamental formulation for the
single quantum spin transition rate with $\Delta m_S = \pm1$.
The anisotropic SAR depends on the local strain fields coupled to defect spins as well as the
spin--strain coupling parameters for $C_{3v}$.
We focus on the ratios of coupling strengths, which are essential for reproducing the
field-direction dependence of SAR.
This is an advantage of our method for analyzing the anisotropic SAR data affected by
various strain-field amplitudes.
The evaluated coupling-strength ratios can be compared straightforwardly with those assessed by
other methods such as first-principles calculation.
\par

This paper is organized as follows.
In Sect.~2, a model Hamiltonian is introduced to describe the spin--strain interaction with the
$C_{3v}$ site symmetry using quadrupole operators (second-rank tensorial forms of spin
operators).
Under the rotation of a strong magnetic field, we investigate the SAR transitions between the two
lowest-lying  $S = 3/2$ levels coupled to local strain fields, which are driven by a standing or
traveling SAW.
In Sect.~3, we demonstrate how to evaluate the spin--strain coupling parameters for $C_{3v}$
from the field-direction dependence of SAR transition rate, considering a deviation of the
parameters from the spherical symmetry as well.
The effect of the SAW propagating in general directions is also discussed.
In particular, we demonstrate how the anisotropic $C_{3v}$ symmetry appears in the
transition rate.
For this purpose, we also study the double quantum spin ($\Delta m_S = \pm 2$) transition rate.
In the last section, we present the conclusion of this paper.
In Appendix~A, we explicitly show the matrix forms of spin and quadrupole operators for
$S = 3/2$.
The unitary transformation of these operators is also given in the case of the rotation of a
magnetic field.
In Appendix~B, we describe a spherically symmetric spin--strain interaction model for comparison.
Appendix~C shows our formulation of the double quantum spin transition rate.

%%%%%%%%%%%%%%%%%%%%%%%%%%%%%%%%%%%%%%%%%%%%%%%%%%%%%%%%%%%%%%%%%%%%%%%%%%%%%%%%%%%%%%%%%%%%%%%%%%%
\section{Model and Formulation}
%%%%%%%%%%%%%%%%%%%%%%%%%%%%%%%%%%%%%%%%%%%%%%%%%%%%%%%%%%%%%%%%%%%%%%%%%%%%%%%%%%%%%%%%%%%%%%%%%%%
\subsection{Spin--strain interaction driven by SAW}
\begin{figure}
\begin{center}
\includegraphics[width=4.5cm,clip]{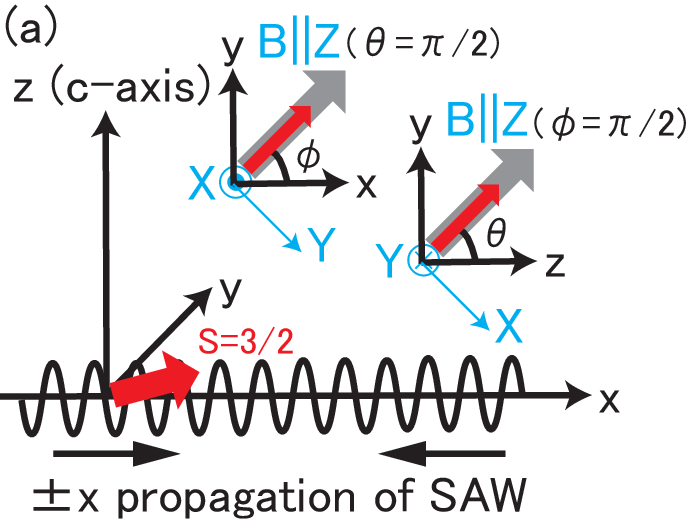}
\includegraphics[width=3.2cm,clip]{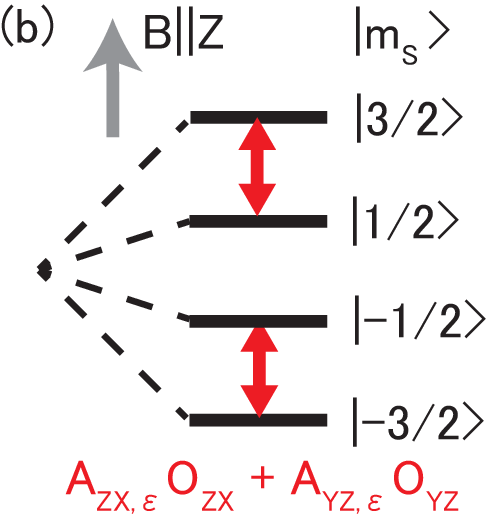}
\end{center}
\caption{
(Color online)
(a) Illustration of $V_{\rm Si}$ defect spin ($S = 3/2$) coupled to the $\pm x$ propagating
SAWs in the SiC surface layer perpendicular to the crystallographic axis $c \parallel z$.
The applied magnetic field $\bB = B (\sin \theta \cos \phi, \sin \theta \sin \phi, \cos \theta)$ is rotated
in the $xy$ ($\theta = \pi /2$) and $yz$ ($\phi = \pi / 2$) planes.
(b) Level splitting for a sufficiently strong magnetic field $B$.
The two levels are coupled via the $ZX$ and $YZ$ quadrupole components in the single quantum
spin transition.
}
\label{fig:1}
\end{figure}

To describe the coupling between the electronic spin and strain components due to elastic
deformations, we study the following form of the spin--strain interaction Hamiltonian:
\begin{align}
H_\varepsilon = \sum_k A_{k, \varepsilon} O_k~~(k = u, v, zx, xy, yz),
\label{eqn:Hep}
\end{align}
where the five quadrupole operators $O_k$ are constructed by the vector of spin operators
$\bS = (S_x, S_y, S_z)$ as
\begin{align}
& O_u = \frac{1}{\sqrt{3}} ( 2 S_z^2 - S_x^2 - S_y^2 ) = \frac{1}{\sqrt{3}} [ 3 S_z^2 - S (S + 1) ],
\nonumber \\
& O_v = S_x^2 - S_y^2,~~O_{zx} = S_z S_x + S_x S_z,
\nonumber \\
& O_{xy} = S_x S_y + S_y S_x,~~~~O_{yz} = S_y S_z + S_z S_y.
\label{eqn:Ok}
\end{align}
The coupling with each quadrupole is expressed by the strain-dependent coupling coefficient
$A_{k, \varepsilon}$.
In the $C_{3v}$ reference frame at the spin site, the spin--strain interaction is characterized by
\cite{Udvarhelyi18a,Udvarhelyi18b}
\begin{align}
& A_{u, \varepsilon} = \frac{ h_a }{4} \varepsilon_u,~~
A_{v, \varepsilon}
= \frac{1}{ 2} \left( \frac{1}{2} h_b \varepsilon_v - h_c \varepsilon_{zx} \right),
\nonumber \\
& A_{zx, \varepsilon}
= \frac{1}{2} \left( - \frac{1}{2} h_{b'} \varepsilon_v + h_{c'} \varepsilon_{zx} \right),~~
A_{xy, \varepsilon} = \frac{1}{2} \left( h_b \varepsilon_{xy} + h_c \varepsilon_{yz} \right),
\nonumber \\
& A_{yz, \varepsilon}
= \frac{1}{2} \left( h_{b'} \varepsilon_{xy} + h_{c'} \varepsilon_{yz} \right),
\label{eqn:Ak}
\end{align}
where $\{ \varepsilon_{\alpha \beta} \}$ ($\alpha, \beta = x, y, z$) are strain components,
$\varepsilon_u = ( 2 \varepsilon_{zz} - \varepsilon_{xx} - \varepsilon_{yy} ) / \sqrt{3}$,
$\varepsilon_v = \varepsilon_{xx} - \varepsilon_{yy}$, and a bulk strain component is disregarded.
There are five independent coupling parameters $h_\lambda$ ($\lambda = a, b, c, b', c'$).
As shown in Fig.~\ref{fig:1}(a), we consider a plane Rayleigh SAW
propagating in either the $+x$ or $-x$ direction, oscillating in the $z$ and $x$ directions (no
displacement along $y$).
In this case, only $A_u$, $A_v$, and $A_{zx}$ are taken into account because the strain
components are restricted to $\varepsilon_{xx}$, $\varepsilon_{zz}$, and $\varepsilon_{zx}$
($\varepsilon_{yy} = \varepsilon_{yz} = \varepsilon_{xy} = 0$) here.
This simplification does not hold when the direction of SAW propagation deviates from
a mirror ($zx$) plane, which will be discussed in Sect.~3.6.
\par
%%%%%%%%%%%%%%%%%%%%%%%%%%%%%%%%%%%%%%
%%%%%%%%%%%%%%%%%%%%%%

\subsection{Spin--strain interaction under rotation of a static magnetic field}
The $S = 3/2$ ground state of $V_{\rm Si}$ is split into the $m_S = \pm 1/2$ and $m_S = \pm 3/2$
Kramers doublets in the $C_{3v}$ crystal-field environment.
In the presence of a magnetic field $\bB$, the local spin Hamiltonian is
$H_B = g \mu_{\rm B} \bS \cdot \bB + D O_u / \sqrt{3}$,
where $g$ is the electron $g$-factor, $\mu_{\rm B}$ is the Bohr magneton, and $2 D$ equals the
zero field splitting.
The field direction is represented by two angles $\theta$ and $\phi$ as
$\bB / B = ( \sin \theta \cos \phi, \sin \theta \sin \phi, \cos \theta )$.
We now consider the case $g \mu_{\rm B} B \gg D$ and neglect the mixing of the $S = 3/2$ quartet
states as shown in Fig.~\ref{fig:1}(b).
\cite{Hernandez-Minguez20}
This allows us to rotate the $\bB$ direction without changing the Zeeman energy shift.
\par

We next introduce a new reference frame $(XYZ)$ where $Z$ is chosen along the $\bB$
direction and define the $Z$-axis unit vector as
$\be_Z = ( \sin \theta \cos \phi, \sin \theta \sin \phi, \cos \theta )$.
For the other orthogonal axes, we define
$\be_X = ( - \cos \theta \cos \phi, - \cos \theta \sin \phi, \sin \theta )$ and
$\be_Y = ( \sin \phi, - \cos \phi, 0 )$.
For the $(XYZ)$ frame, we transform the spin operators from $\bS = ( S_x, S_y, S_z )$ to
$\bS_B = ( S_X, S_Y, S_Z)$ as
$U^\dagger ( \be_\lambda \cdot \bS ) U = S_\lambda~~( \lambda = X, Y, Z )$,
using the unitary matrix $U$ for spin-3/2 in Appendix~A.
For an electronic spin coupled to elastic strains, the quantum spin transitions between the
$S = 3/2$ states are described by the quadrupole operators $O_k$ in Eq.~(\ref{eqn:Ok}).
Under a magnetic field, it is convenient to use the quadrupole operators in the $(XYZ)$ frame.
By the unitary transformation, we transform Eq.~(\ref{eqn:Hep}) to
\begin{align}
H_{\varepsilon, B} = U^\dagger H_{\varepsilon} U \equiv \sum_K A_{K, \varepsilon} O_K,
\label{eqn:HepB}
\end{align}
where the quadrupole operators $O_K$ ($K = U, V, ZX, XY, YZ$) are constructed by the
components of $\bS_B$.
For the finite strain components $\varepsilon_{xx}$, $\varepsilon_{zz}$, and $\varepsilon_{zx}$ of
the SAW considered here, the field-direction-dependent coupling coefficients $A_{K, \varepsilon}$
are given by linear combinations of $A_{u, \varepsilon}$, $A_{v, \varepsilon}$, and
$A_{zx, \varepsilon}$ in Eq.~(\ref{eqn:Ak}).
For the single quantum spin transition with $\Delta m_S = \pm 1$ in Fig.~\ref{fig:1}(b), they
are represented as
\begin{align}
& A_{ZX, \varepsilon} = A_{u, \varepsilon} \cdot \sqrt{3} \sin \theta \cos \theta
+ A_{v, \varepsilon} ( - \sin \theta \cos \theta ) \cos 2 \phi
\nonumber \\
&~~~~~~~~~~~~
+ A_{zx,\varepsilon} ( - \cos 2 \theta ) \cos \phi,
\nonumber \\
& A_{YZ, \varepsilon} = A_{v, \varepsilon} \sin \theta \sin 2 \phi
+ A_{zx, \varepsilon} \cos \theta \sin \phi,
\label{eqn:AK2}
\end{align}
whereas $A_U$, $A_V$, and $A_{ZX}$ are not involved in Eq.~(\ref{eqn:HepB}).
Hence, the spin--strain interaction is described by $H = B S_Z + H_{\varepsilon, B}$ in the
presence of a magnetic field.
\par

\subsection{Single quantum spin transition driven by SAW}
For the strain components driven by a SAW, we adopt a realistic assumption that
$\varepsilon_{xx}$ and $\varepsilon_{zz}$ are real, whereas
$\varepsilon_{zx} \equiv - i \varepsilon''_{zx}$ is purely imaginary ($\varepsilon''_{zx}$ is real).
\cite{Hernandez-Minguez20}
Accordingly, $A_u$ is real and the other components are represented by
$A_v = | A_v | e^{ i \theta_v }$ and $A_{zx} = | A_{zx} | e^{ i \theta_{zx} }$,
where the subscripts $\varepsilon$ of $A_{k, \varepsilon}$ are omitted.
From this representation, the coupling parameters in Eq.~(\ref{eqn:Ak}) are also related to $A_v$ and $A_{zx}$ as
\begin{align}
& \frac{ h_b }{4} \varepsilon_{xx} = | A_v | \cos \theta_v,~~
\frac{ h_c }{2} \varepsilon''_{zx} = | A_v | \sin \theta_v,
\nonumber \\
& - \frac{ h_{b'} }{4} \varepsilon_{xx} = | A_{zx} | \cos \theta_{zx},~~
- \frac{ h_{c'} }{2} \varepsilon''_{zx} = | A_{zx} | \sin \theta_{zx}.
\label{eqn:hbc}
\end{align}
In a spherical approximation, $h_a = h_b = h_{c'}$ and $h_c = h_{b'} = 0$;
the latter indicates $\theta_v = 0$ and $| \theta_{zx} | = \pi / 2$.
As noted in Appendix~B, we have $\tan \theta_v / \tan \theta_{zx} = ( h_c h_{b'} ) / ( h_b h_{c'} )$,
which represents a deviation from the spherical symmetry.
In general, $\theta_v$ and $\theta_{zx}$ change satisfying $0 \le | \theta_v |, | \theta_{zx} | \le \pi$.
\par

In Fig.~\ref{fig:1}(b), we restrict ourselves to the two lowest-lying states
$| {\rm g} \rangle = | m_S = -3/2 \rangle$ and $| {\rm e} \rangle = | m_S = -1/2 \rangle$ to calculate
the transition matrix element
$M = \langle {\rm e} | A_{ZX, \varepsilon} O_{ZX} + A_{YZ, \varepsilon} O_{YZ} | {\rm g} \rangle$.
In SAR measurements, V$_{\rm Si}$ centers can be coupled
simultaneously to two SAWs propagating along the $+ x$ and $- x$ directions with intensities
$I_+$ and $I_-$, respectively.
The transition rate between the $| {\rm g} \rangle$ and $| {\rm e} \rangle$ states
is proportional to the sum of the SAW contributions,
\begin{align}
W \propto \left\langle | M_+ |^2 \frac{ I_+ }{ I_+ + I_- } + | M_- |^2 \frac{ I_- }{ I_+ + I_- }
\right\rangle,
\label{eqn:WM}
\end{align}
where the transition matrix elements $M_+$ and $M_-$ are related to $+ x$ and $- x$
propagations, respectively, and $M_-$ is given by replacing
$\varepsilon''_{zx} \rightarrow - \varepsilon''_{zx}$ ($\theta_v \rightarrow - \theta_v$ and
$\theta_{zx} \rightarrow - \theta_{zx}$) in $M_+$.
\cite{Hernandez-Minguez20}
The bracket $\langle \cdots \rangle$ indicates the average of the strain amplitudes along $x$
(surface layer) and $z$ (depth) considering the spatial distribution of V$_{\rm Si}$ centers.
\par

%%%%%%%%%%%%%%%%%%%%%%%%%%%%%%%%%%%%%%%%%%%%%%%%%%%%%%%%%%%%%%%%%%%%%%%%%%%%%%%%%%%%%%%%%%%%%%%%%%%
\section{Results}
%%%%%%%%%%%%%%%%%%%%%%%%%%%%%%%%%%%%%%%%%%%%%%%%%%%%%%%%%%%%%%%%%%%%%%%%%%%%%%%%%%%%%%%%%%%%%%%%%%%
\subsection{Field-direction-dependent transition rate}
First, we consider the transition matrix element $M_{xy, \pm}$
for the inplane ($xy$) rotation of $\bB = B (\cos \phi, \sin \phi, 0)$, where $\theta = \pi / 2$.
Using Eq.~(\ref{eqn:AK2}), we obtain
\begin{align}
M_{xy, \pm} = \sqrt{3} | A_v | [ - | a_{zx} | e^{ \pm i ( \theta_{zx} - \theta_v )} \cos \phi + i \sin 2 \phi ],
\end{align}
with $a_{zx} = A_{zx} / |A_v|$.
The transition rate is calculated as
\begin{align}
& W_{xy} \propto \Big\langle 3 | A_v |^2 [ | a_{zx} |^2 \cos^2 \phi + \sin^2 2 \phi
\nonumber \\
&~~~~~~~~~~~~
- 2 \eta | a_{zx} | \sin ( \theta_{zx} - \theta_v ) \cos \phi \sin 2 \phi ] \Big\rangle,
\label{eqn:Wxy}
\end{align}
where the parameter $\eta = ( I_+ - I_- ) / ( I_+ + I_- )$ represents a standing wave for $\eta = 0$
and a one-way traveling wave for $\eta = \pm 1$.
The intermediate case $0 < | \eta | < 1$ represents different weights of combined
counterpropagating SAWs.
The last term in Eq.~(\ref{eqn:Wxy}) indicates the difference between the $C_{3v}$ symmetry and
the spherical approximation.
This enables us to evaluate the coupling parameters in the spin--strain interaction from
experimental results as discussed below.
\par

The values of $\langle | a_{zx} | \rangle$ and $\Delta \theta \equiv \theta_{zx} - \theta_{zx}$ can be
determined from $W_{xy}$ as a function of $\phi$.
From these values, we can evaluate the coupling-strength ratios $h_{b'} / h_b$ and
$h_{c'} / h_c$ in Eq.~(\ref{eqn:hbc}).
In this study, we pay attention to the maximum transition rate $W_{xy}$ for $\eta = 0$
and $| \eta | = 1$, assuming that the combination of $\pm x$ propagating SAWs can be
controlled by adjusting the edge reflection. 
\par

\subsection{Transition rate $W_{xy}$ for $\eta = 0$}
\begin{figure}
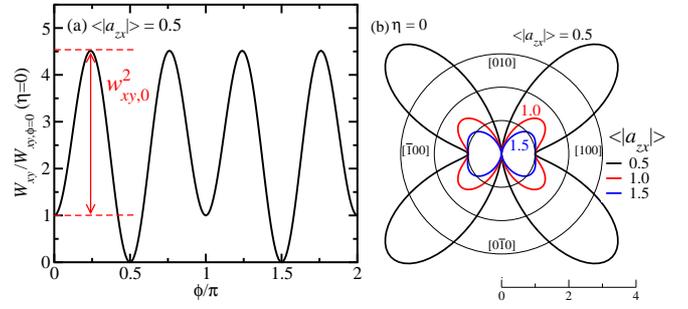

\begin{center}
\includegraphics[width=4.6cm,clip]{fig2a.eps}
\includegraphics[width=3.9cm,clip]{fig2b.eps}
\end{center}
\caption{
(Color online)
(a) Transition rate for the standing SAW ($\eta = 0$) plotted as a function of the
field-rotation angle $\phi$ in the $xy$ plane for $\langle | a_{zx} | \rangle = 0.5$.
A measurable value of $w_{xy,0}^2$ indicated by the two-headed arrow is used for calculating
$\langle | a_{zx} | \rangle$.
(b) Polar representation of the normalized rate plotted as the radius coordinate with respect to
$\phi$ for $\langle | a_{zx} | \rangle = 0.5, 1.0$, and $1.5$.
}
\label{fig:2}
\end{figure}

In Eq.~(\ref{eqn:Wxy}), we consider $W_{xy}$ as a function of $\langle | a_{zx} | \rangle$ to simplify
the calculation of the maximums, assuming that $\langle | a_{zx} | \rangle$ is substituted for
$\sqrt{ \langle | a_{zx} |^2 \rangle }$.
For $\eta = 0$, the maximum $W_{xy, {\rm max}}$ is obtained for
$\phi = (1/2) \arccos ( \langle | a_{zx} |^2 \rangle / 4 )$.
We then introduce the quantity
\begin{align}
w_{xy, 0}^2 \equiv
\left( \frac{ W_{xy, {\rm max}} - W_{xy, \phi = 0} }{ W_{xy, \phi = 0} } \right)_{ \eta = 0 }
= \frac{1}{ \langle | a_{zx} |^2 \rangle} \left( 1 - \frac{ \langle | a_{zx} |^2 \rangle }{4} \right)^2
\label{eqn:wxy0}
\end{align}
and solve
\begin{align}
\langle | a_{zx} | \rangle = 2 \left( - w_{xy, 0} + \sqrt{ 1 + w_{xy, 0}^2 } \right).
\label{eqn:azxwxy}
\end{align}
In Fig.~\ref{fig:2}(a), $W_{xy}$ is plotted for
$\langle | a_{zx} | \rangle = \sqrt{ \langle | a_{zx} |^2 \rangle } = 0.5$.
A measurable value of $w_{xy,0}^2$ determines $\langle | a_{zx} | \rangle$ in Eq.~(\ref{eqn:azxwxy}).
Figure~\ref{fig:2}(b) shows the polar representation of $W_{xy} / W_{xy, \phi = 0}$ as a function of
$\phi$ for various values of $\langle | a_{zx} | \rangle$.
The contours of the plots are symmetric with respect to the inversion of $B_x$
(for instance, $\phi \rightarrow \pi - \phi$) owing to a mirror plane of $zx$ and the inversion of
$B_y$ 
($\phi \rightarrow 2 \pi - \phi$) owing to the time-reversal symmetry for $\eta = 0$.
\cite{Hernandez-Minguez20}
For small values of $\langle | a_{zx} | \rangle < 1$, $W_{xy}$ shows four peaks at
$\phi / \pi \simeq 1/4, 3/4, 5/4$, and $7/4$, which are markedly increased by a large deviation from
the spherical symmetry at $\langle | a_{zx} | \rangle \simeq 1$ in the spin--strain interaction.
\par

%%%%%%%%%%%%%%%%%%%%%%%%%%%%%%%%%%%%%%
%%%%%%%%%%%%%%%%%%%%%%
For the spherical model in Eq.~(\ref{eqn:Hep-s}),
$\langle | a_{zx} | \rangle$ only depends on
$2 \langle | \varepsilon''_{zx} \varepsilon_{xx} | \rangle / \langle \varepsilon_{xx}^2 \rangle$.
According to an estimate for the SAR measurements,
\cite{Hernandez-Minguez20}
this value approximately equals 1.2.
For $\langle | a_{zx} | \rangle > 1$, $W_{xy}$ does not exhibit the V-shaped angular dependence
at $\phi = 0$ and $\pi$ as shown in Fig.~\ref{fig:2}(b), whereas the V-shaped dependence is
found in the SAR data.
\cite{Hernandez-Minguez20}
This discrepancy can be solved by considering a deviation from the spherical symmetry
$| h_{c'} / h_b | < 1$ for $C_{3v}$.
Indeed, we can estimate $w_{xy,0}^2 \simeq 2$ from the SAR data and obtain
$\langle | a_{zx} | \rangle \simeq 0.6$ from Eq.~(\ref{eqn:azxwxy}).

\subsection{Transition rate $W_{xy}$ for $| \eta | = 1$}
\begin{figure}
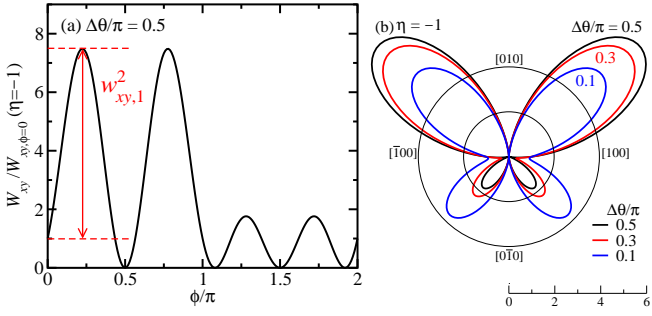

\begin{center}
\includegraphics[width=4.7cm,clip]{fig3a.eps}
\includegraphics[width=3.7cm,clip]{fig3b.eps}
\end{center}
\caption{
(Color online)
(a) Transition rate for the $-x$ propagating SAW ($\eta = - 1$) plotted as a function of $\phi$  for
$\Delta \theta \equiv \theta_{zx} - \theta_v = 0.5 \pi$.
Here,  we fix $\langle | a_{zx} | \rangle = 0.5$.
The measurable value $w_{xy,1}^2$ indicated by the two-headed arrow is used for evaluating
$\theta_v$ and $\theta_{zx}$.
(b) Polar representation of the normalized transition rate for $\Delta \theta / \pi = 0.5, 0.3$, and
$0.1$.
}
\label{fig:3}
\end{figure}

The above procedure is also applied to $W_{xy}$ for $|\eta| = 1$.
Similarly, we define
\begin{align}
& w_{xy, 1}^2 \equiv
\left( \frac{ W_{xy, {\rm max}} - W_{xy, \phi = 0} }{ W_{xy, \phi = 0} } \right)_{ |\eta| = 1 }
\nonumber \\
&~~~~~~
=  \langle | a_{zx} |^2 \rangle^{-1}
\left(  \langle | a_{zx} |^2 \rangle \cos^2 \phi_{\rm max} + \sin^2 2 \phi_{\rm max} \right.
\nonumber \\
&~~~~~~~~~~~~~~\left.
+ 2 \gamma_{xy}  \langle | a_{zx} | \rangle \cos \phi_{\rm max} \sin 2 \phi_{\rm max} \right) -1
\nonumber \\
&~~~~~~
( \gamma_{xy} \equiv | \sin ( \theta_{zx} - \theta_v ) | ).
\label{eqn:wxy1}
\end{align}
Here, $\phi_{\rm max}$ is the angle for a maximum $W_{xy, {\rm max}}$.
Using the value of $\langle | a_{zx} | \rangle$ evaluated from Eq.~(\ref{eqn:azxwxy}), we obtain the
parameter $\gamma_{xy}$ from the measurable values of $w_{xy,1}$ and $\phi_{\rm max}$.
In Fig.~\ref{fig:3}(a), $W_{xy}$ is plotted for $\eta = -1$, where
$\Delta \theta \equiv \theta_{zx} - \theta_v = 0.5 \pi$ is fixed.
The value of $\gamma_{xy} \equiv | \sin \Delta \theta |$ can be evaluated from
Eq.~(\ref{eqn:wxy1}) with the measurable values of $w_{xy,1}^2$ and $\phi_{\rm max}$.
Figure~\ref{fig:3}(b) shows the polar plots of $W_{xy} / W_{xy, \phi = 0}$ for various values of
$\Delta \theta$.
The asymmetric contours of the plots ($W_{xy, \phi} \ne W_{xy,- \phi}$) are due to the
broken time-reversal symmetry for $\eta \ne 0$.
Note that the maximum transition rate decreases with decreasing in $\Delta \theta / \pi$
from $0.5$.
The symmetric $\phi$ dependence in Fig.~\ref{fig:2}(b) is restored at $\Delta \theta = 0$ even for
$\eta \ne 0$ owing to the disappearance of the last term in Eq.~(\ref{eqn:Wxy}).
It is the special case $\theta_v = \theta_{zx}$ for $C_{3v}$ where the coupling parameters satisfy
$h_{b'} / h_b = h_{c'} / h_c$ in Eq.~(\ref{eqn:hbc}).
\par

%%%%%%%%%%%%%%%%%%%%%%%%%%%%%%%%%%%%%%
%%%%%%%%%%%%%%%%%%%%%%
For the spherical model in Eq.~(\ref{eqn:Hep-s}), the $\phi$ dependence for $| \eta | = 1$
corresponds to $\Delta \theta / \pi = 0.5$ because of $\theta_v = 0$ ($h_c = 0$) and
$\theta_{zx} / \pi = 0.5$ ($h_{b'} = 0$) as shown in Fig.~\ref{fig:b1}(b).
For a large deviation $| \Delta \theta / \pi | \ll 0.5$ from the spherical symmetry, the $\phi$
dependence becomes symmetric as shown for $\eta = 0$ in Fig.~\ref{fig:2}(b).
\par

\subsection{Transition rate under inplane (yz) field rotation}
To determine $\theta_v$ and $\theta_{zx}$, we apply a similar formulation to a
different spin transition rate $W_{yz}$ for the inplane ($yz$) rotation of
$\bB = B ( 0, \sin \theta, \cos \theta )$, where $\phi = \pi / 2$.
As given by Eqs.~(\ref{eqn:wxy0}) and (\ref{eqn:wxy1}), we can express
\begin{align}
& w_{yz, 0}^2 \equiv
\left( \frac{ W_{yz, {\rm max}} - W_{yz, \theta = 0} }{ W_{yz, \theta = 0} } \right)_{ \eta = 0 }
= \frac{ \langle \alpha_u^2 \rangle }{ \langle 4 | a_{zx} |^2 \rangle }
\left( 1 - \frac{ \langle | a_{zx} |^2 \rangle }{ \langle \alpha_u^2 \rangle } \right)^2
\nonumber \\
&~~~~~~
( \alpha_u^2 \equiv 3 a_u^2 + 2 \sqrt{3} a_u \cos \theta_v + 1,~a_u = A_u / | A_v |),
\label{eqn:wyz0}
\end{align}
and
\begin{align}
& w_{yz, 1}^2 \equiv
\left( \frac{ W_{yz, {\rm max}} - W_{yz, \theta = 0} }{ W_{yz, \theta = 0} } \right)_{ |\eta| = 1 }
\nonumber \\
&~~~~~~
=  \langle | a_{zx} |^2 \rangle^{-1}
\left(  \langle | a_{zx} |^2 \rangle \cos^2 \theta_{\rm max}
+ \frac{ \langle \alpha_u^2 \rangle }{4} \sin^2 2 \theta_{\rm max} \right.
\nonumber \\
&~~~~~~~~~~~~~~\left.
+ 2 \gamma_{yz}  \langle | a_{zx} | \rangle \cos \theta_{\rm max} \sin 2 \theta_{\rm max} \right) -1
\nonumber \\
&~~~~~~
( 2 \gamma_{yz} \equiv | \sqrt{3} \langle a_u \rangle \sin \theta_{zx}
+ \sin ( \theta_{zx} - \theta_v ) | ).
\label{eqn:wyz1}
\end{align}
Here, $\theta_{\rm max}$ is the angle for a maximum $W_{yz, {\rm max}}$ ($| \eta | = 1$).
From Eq.~(\ref{eqn:wyz0}), we obtain
\begin{align}
\langle \alpha_u^2 \rangle
= \langle | a_{zx} |^2 \rangle \left( - w_{yz,0} + \sqrt{ 1 + w_{yz,0}^2 } \right)^{-2},
\end{align}
with the measurable value $w_{yz,0}$ and the $\langle | a_{zx} | \rangle$ determined using
Eq.~(\ref{eqn:wxy0}). 
The parameter $\gamma_{yz}$ can be determined from the measurable values of $w_{yz,1}$ and
$\theta_{\rm max}$, where $\langle | a_{zx} | \rangle$ and $\langle \alpha_u^2 \rangle$ are
already known.
In Table~I,  we summarize the measurable and evaluated quantities for the coupling
parameters in the spin--strain interaction.
The parameters $\langle | a_{zx} | \rangle$, $\langle \alpha_u^2 \rangle$, $\gamma_{xy}$, and
$\gamma_{yz}$ are used for evaluating the coupling-strength ratios $h_{b'} / h_b$ and
$h_{c'} / h_c$ in the spin--strain interaction for $C_{3v}$, which will be discussed in the next
subsection.
\par
%%%%%%%%%%%%%%%%%%%%%%%%%%%%%%%%%%%%%%
\begin{table}
\caption{
List of measurable and calculated quantities for the evaluation of the coupling
parameters.
The values of $w_{xy,0}$, $w_{xy,1}$, and $\phi_{\rm max}$ are obtained from the $\phi$
dependence of the single quantum spin transition rate $W_{xy}$ in the $xy$ plane, and those of
$w_{yz,0}$, $w_{yz,1}$, and $\theta_{\rm max}$ are obtained from the $\theta$ dependence of
$W_{yz}$ in the $yz$ plane.
The values of $\langle | a_{zx} | \rangle$, $\gamma_{xy}$, $\langle \alpha_u^2 \rangle$,
and $\gamma_{yz}$ in the second column are calculated from the measurable
values listed in the first column.
In the third column, $| \eta | = 0$ and $| \eta | = 1$ represent the standing and traveling SAWs
propagating along the $x$-axis, respectively.
}
\begin{center}
\begin{tabular}{lll} \hline
Measurable & Calculated & $| \eta |$ (SAW) \\ \hline
$w_{xy,0}$ & $\langle | a_{zx} | \rangle$ & $0$ \\
$w_{xy,1}$, $\phi_{\rm max}$ & $\gamma_{xy}$ using $\langle | a_{zx} | \rangle$ & $1$ \\
$w_{yz,0}$ & $\langle \alpha_u^2 \rangle$ using $\langle | a_{zx} | \rangle$ & $0$ \\
$w_{yz,1}$, $\theta_{\rm max}$ & $\gamma_{yz}$ using $\langle | a_{zx} | \rangle$ and
$\langle \alpha_u^2 \rangle$ & $1$ \\ \hline
\end{tabular}
\end{center}
\end{table}
%%%%%%%%%%%%%%%%%%%%%%

\subsection{Evaluation of coupling-strength ratios of spin--strain interaction}
\begin{figure}
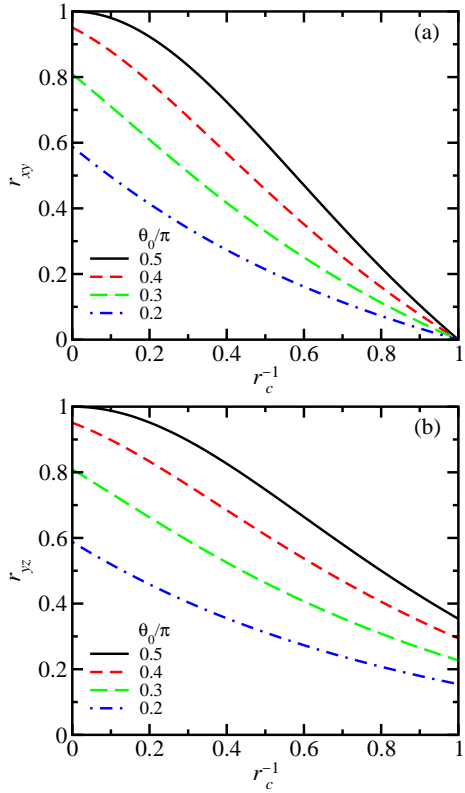

\begin{center}
%\includegraphics[width=6cm,clip]{s32rzth02t05figa.eps}
%\hspace*{0.5cm}
%\includegraphics[width=6cm,clip]{s32rxth02t05figb.eps}
\includegraphics[width=6cm,clip]{fig4a.eps}
\includegraphics[width=6cm,clip]{fig4b.eps}
\end{center}
\caption{
(Color online)
(a) Plots of the parameter $\gamma_{xy}$ as a function of
$r_c^{-1} \equiv \sin \theta_v / \sin \theta_{zx}$ for various values of
$\theta_0 = \theta_v + \theta_{zx}$, related to the transition rate $W_{xy}$ under the rotation of a
magnetic field in the $xy$ plane.
(b) Plots of $\gamma_{yz}$ as a function of $r_c^{-1}$, where $\sqrt{3} \langle a_u \rangle = 1$ is
used, related to $W_{yz}$ under the rotation of a magnetic field in the $yz$ plane.
The spherical spin--strain coupling case corresponds to $\gamma_{xy} = \gamma_{yz} = 1$ at
$r_c^{-1} = 0$.
}
\label{fig:4}
\end{figure}

Now, $\gamma_{xy}$ and $\gamma_{yz}$ are determined from the measurable values.
From Eq.~(\ref{eqn:hbc}), it is convenient to introduce the following parameters:
\begin{align}
r_b \equiv \frac{ \cos \theta_{zx} }{ \cos \theta_v } = - \frac{ h_{b'} }{ h_b } \langle | a_{zx} | \rangle,~~
r_c \equiv \frac{ \sin \theta_{zx} }{ \sin \theta_v } = - \frac{ h_{c'} }{ h_c } \langle | a_{zx} | \rangle,
\label{eqn:rbc}
\end{align}
which are given by Eq.~(\ref{eqn:hbc}).
To evaluate the coupling-strength ratios $h_{b'} / h_b$ and $h_{c'} / h_c$, we determine
$r_b$ and $r_c$ from the values of $\gamma_{xy}$ and $\gamma_{yz}$ as follows.
Here, let us change $\theta_v$ and $\theta_{zx}$ under the condition
$\theta_v + \theta_{zx} = \theta_0$ ($0 \le \theta_v < \theta_0 / 2 \le \pi / 4$), and rewrite
$r_b$ and $r_c$ as
\begin{align}
& r_b \equiv \frac{ \cos \theta_{zx} }{ \cos \theta_v }
= \cos \theta_0 + \sin \theta_0 \tan \theta_v~~( \cos \theta_0 \le r_b \le 1 ),
\label{eqn:rb2} \\
& r_c \equiv \frac{ \sin \theta_{zx} }{ \sin \theta_v }
= - \cos \theta_0 + \frac{ \sin \theta_0 }{ \tan \theta_v }~~( r_c \ge 1).
\label{eqn:rc2}
\end{align}
Eliminating $\tan \theta_v$, we obtain
\begin{align}
r_b = \frac{ r_c^{-1} + \cos \theta_0 }{ 1 + r_c^{-1} \cos \theta_0 }~~(0 \le r_c^{-1} \le 1).
\label{eqn:rbc2}
\end{align}
On the other hand, $r_b$ and $r_c$ satisfy $r_b^2 \cos^2 \theta_v + r_c^2 \sin^2 \theta_v = 1$,
which leads to
\begin{align}
& \sin^2 \theta_v = \frac{ 1 - r_b^2 }{ r_c^2 - r_b^2 }
= r_c^{-2} \left[ 1 + \frac{ r_b^2 ( r_c^2 - 1 ) }{ r_c^2 ( 1 - r_b^2 ) } \right]^{-1},~~
 \cos^2 \theta_v = \frac{ r_c^2 - 1 }{ r_c^2 - r_b^2 },
\nonumber \\
& \sin^2 ( \theta_{zx} - \theta_v )
= ( \sin \theta_{zx} \cos \theta_v - \cos \theta_{zx} \sin \theta_v )^2
\nonumber \\
&~~~~~~~~~~~~~~~~~~~~~
= ( r_c \sin \theta_v \cos \theta_v - r_b \cos \theta_v \sin \theta_v )^2
\nonumber \\
&~~~~~~~~~~~~~~~~~~~~~
= (r_c - r_b)^2 \sin^2 \theta_v \cos^2 \theta_v
= \frac{ (1 - r_b^2)(r_c^2 - 1) }{ (r_b + r_c)^2 }
\nonumber \\
&~~~~~~~~~~~~~~~~~~~~~
= \left[ 1 + \frac{ ( 1 + r_b r_c )^2 }{ ( 1 - r_b^2 )( r_c^2 - 1) } \right]^{-1}.
\label{eqn:sin2zxv}
\end{align}
Combining Eqs.~(\ref{eqn:rbc2}) and (\ref{eqn:sin2zxv}), we obtain
\begin{align}
\gamma_{xy} = \frac{ ( 1 - r_c^{-2} ) \sin \theta_0 }
{ \{ (1 - r_c^{-2} )^2 \sin^2 \theta_0 + [ 2 r_c^{-1} + (1 + r_c^{-2}) \cos \theta_0 ]^2 \}^{1/2} }
\label{eqn:gxy}
\end{align}
and
\begin{align}
\gamma_{yz} = \frac{1}{2} \left| \frac{ \sqrt{3} \langle a_u \rangle \sin \theta_0 }
{ [ (r_c^{-1} + \cos \theta_0)^2 + \sin^2 \theta_0 ]^{1/2} } + \gamma_{xy} \right|,
\label{eqn:gyz}
\end{align}
as a function of $r_c^{-1}$ ($0 \le r_c^{-1} \le 1$).
Here, $\langle a_u \rangle$ is related to $\langle \alpha_u^2 \rangle$ in Eq.~(\ref{eqn:wyz0}),
and the latter is calculated from the measurable value of $w_{yz,0}$.
In Figs.~\ref{fig:4}(a) and 4(b), we show the $r_c^{-1}$ dependences of $\gamma_{xy}$ and
$\gamma_{yz}$, respectively.
We can extract $r_c^{-1}$ and $\theta_0$ to reproduce the observed values of
$\gamma_{xy}$ and $\gamma_{yz}$ shown in Table~I.
As a result, we obtain $r_b$ using Eq.~(\ref{eqn:rbc2}).
Finally, we can determine $h_{b'} / h_b$ and $h_{c'} / h_c$ using $r_b$ and $r_c$ with the
value of $\langle | a_{zx} | \rangle$ in Eq.~(\ref{eqn:rbc}).
\par

%%%%%%%%%%%%%%%%%%%%%%
%%%%%%%%%%%%%%%%%%%%%%
For the NV centers in diamond, the coupling strengths $| h_{b'} |$ and $| h_{c'} |$ are related to the
$zx$ and $yz$ quadrupole components as shown in Eq.~(\ref{eqn:Ak}).
They have been considered negligibly small in earlier studies,
\cite{Barry20,Kehayias19}
whereas it was reported that $h_{b'} / h_b \simeq - 0.46$ and $h_{c'} / h_c \simeq - 0.14$ were
assessed by the first-principles calculation.
\cite{Udvarhelyi18a}
In addition, recent acoustic measurements indicated a relatively large spin--stress coupling
strength related to $h_{b'}$ and $h_{c'}$.
\cite{Chen20}
It does not seem so difficult to probe the quantum spin transitions driven by the $zx$ and $yz$
quadrupole components; however, this has not been established for the NV centers yet.

\subsection{SAW propagating in general directions}
Here, we demonstrate how $W_{xy}$ is affected by the SAW propagation direction deviating from
the crystallographic axis.
We choose the $x'$-axis along the SAW propagation direction and introduce a new reference
frame $(x'y'z)$ obtained by the transformation
\begin{align}
\left(
\begin{array}{c}
x' \\
y'
\end{array}
\right)
= \left(
\begin{array}{cc}
\cos \varphi & \sin \varphi \\
- \sin \varphi & \cos \varphi
\end{array}
\right)
\left(
\begin{array}{c}
x \\
y
\end{array}
\right).
\end{align}
Here, the $x'$-axis is rotated from $x$ by angle $\varphi$.
As in the $\varphi = 0$ case, the finite strain components are restricted to $\varepsilon_{x'x'}$ and
$\varepsilon_{x'z}$ for the SAWs (no displacement along $y'$).
In Eq.~(\ref{eqn:Ak}), the strain tensor components
$\{ \varepsilon_{xx}, \varepsilon_{yy}, \varepsilon_{xy}, \varepsilon_{zx}, \varepsilon_{yz} \}$ are
replaced as follows:
\begin{align}
& \varepsilon_{xx} \rightarrow \varepsilon_{x' x'} \cos^2 \varphi,~~
\varepsilon_{yy} \rightarrow \varepsilon_{x' x'} \sin^2 \varphi,~~
\varepsilon_{xy} \rightarrow \varepsilon_{x'x'} \sin \varphi \cos \varphi,
\nonumber \\
& \varepsilon_{zx} \rightarrow \varepsilon_{x' z} \cos \varphi,~~
\varepsilon_{yz} \rightarrow \varepsilon_{x' z} \sin \varphi.
\label{eqn:newstrain}
\end{align}
In the $(XYZ)$ frame for a finite $\bB$, we use the unitary transformation in Eq.~(\ref{eqn:OkOK}) for
$\theta = \pi / 2$ and obtain the field-direction-dependent coupling coefficients,
\begin{align}
& A_{ZX, \varepsilon} = \frac{1}{2} \left[ h_{c'} \varepsilon_{x' z} \cos \varphi
- \frac{1}{2} h_{b'} \varepsilon_{x' x'} ( \cos^2 \varphi - \sin^2 \varphi ) \right] \cos \phi
\nonumber \\
&~~~~~~~~~~
+ \frac{1}{2} \left[ h_{c'} \varepsilon_{x' z} \sin \varphi
+ h_{b'} \varepsilon_{x' x'} \sin \varphi \cos \varphi \right] \sin \phi, \\
& A_{YZ, \varepsilon} = \frac{1}{2} \left[ - h_c \varepsilon_{x' z} \cos \varphi
+ \frac{1}{2} h_b \varepsilon_{x' x'} ( \cos^2 \varphi - \sin^2 \varphi ) \right] \sin 2 \phi
\nonumber \\
&~~~~~~~~~~
- \frac{1}{2} \left[ h_c \varepsilon_{x' z} \sin \varphi
+ h_b \varepsilon_{x' x'} \sin \varphi \cos \varphi \right] \cos 2 \phi,
\end{align}
for the $ZX$ and $YZ$ quadrupole components, respectively.
We obtain the transition matrix elements
$M_\pm = \langle {\rm e} | A_{ZX, \varepsilon} O_{ZX}
+ A_{YZ, \varepsilon} O_{YZ} | {\rm g} \rangle$
for $\varepsilon_{x' z} = \mp i \varepsilon''_{x'z}$ ($\varepsilon_{x' z}$ is purely imaginary):
\begin{align}
& M_{xy, \pm} = \frac{ \sqrt{3} }{ 4 }
\{ [ h_{b'} \varepsilon_{x' x'} \cos ( \phi + 2 \varphi )
\mp 2 h_c \varepsilon''_{x' z} \sin ( 2 \phi + \varphi ) ]
\nonumber \\
&~~~~~~~~~~~~~~~~
+ i [ h_b \varepsilon_{x' x'} \sin 2 ( \phi - \varphi )
\pm 2 h_{c'} \varepsilon''_{x' z} \cos ( \phi - \varphi ) ] \}.
\end{align}
After calculating $| M_{xy, \pm} |^2$, we obtain the transition rate
\begin{align}
& W_{xy,0} \propto
\langle \varepsilon_{x' x'}^2 [ h_b^2 \sin^2 2 \phi' + h_{b'}^2 \cos^2 (\phi' + 3 \varphi) ]
\nonumber \\
&~~~~~~~~
+ 4 \varepsilon_{x' z}''^2 [ h_{c'}^2 \cos^2 \phi' + h_c^2 \sin^2 (2 \phi' + 3 \varphi) ] \rangle
\label{eqn:Wc3v}
\end{align}
for $\eta = 0$.
Here, $\phi'$ ($= \phi - \varphi$) is the field-rotation angle measured from the $x'$-axis.
\par

To reveal the $C_{3v}$ symmetry inherent in $W_{xy,0}$, we simplify Eq.~(\ref{eqn:Wc3v}) as
\begin{align}
& W_{xy,0} \propto
2 ( 1 + \lambda ) + ( 1 + \lambda \cos 6 \varphi )( \cos 2 \phi' - \cos 4 \phi' )
\nonumber \\
&~~~~~~~~
+ \lambda \sin 6 \varphi ( \sin 4 \phi' - \sin 2 \phi' ),
\label{eqn:Wc3vs}
\end{align}
assuming that
$\langle | h_{b'} \varepsilon_{x' x'} |^2 \rangle
= \langle | 2 h_c \varepsilon''_{x' z} |^2 \rangle
= \lambda \langle | h_b \varepsilon_{x' x'} |^2 \rangle
= \lambda \langle | 2 h_{c'} \varepsilon''_{x' z} |^2 \rangle$.
Here, the parameter $\lambda$ represents a deviation from the spherical symmetric form of the
spin--strain interaction.
%%%%%%%%%%%%%%%%%%%%%%
\begin{figure}
\begin{center}
\includegraphics[width=4cm,clip]{fig5left.eps}
\includegraphics[width=3cm,clip]{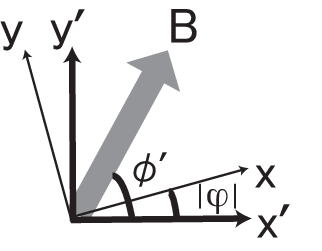}
\end{center}
\caption{
(Color online)
Polar representation of normalized transition rate $W_{xy, 0} / W_{xy, 0, \phi' = 0}$ for $\eta = 0$
plotted as a function of $\phi'$ for $\lambda = 0.0, 1.0$, and $5.0$ in Eq.~(\ref{eqn:Wc3vs}), where
$\varphi = - \pi / 12$ is used.
The angle $\phi'$ is the $\bB$ direction measured from the SAW propagation axis ($x'$-axis).
}
\label{fig:5}
\end{figure}
%%%%%%%%%%%%%%%%%%%%%%
The last term in Eq.~(\ref{eqn:Wc3vs}) describes the odd function nature of $\phi'$.
For $\lambda \ne 0$, this causes an asymmetric $\phi'$ dependence with respect to the $x'$- and
$y'$-axes.
The amplitude of the asymmetry is proportional to $\sin 6 \varphi$ and shows a periodic change
with respect to each $\pi / 3$, reflecting the $C_{3v}$ symmetry.
It vanishes for $\varphi =0, \pm \pi / 3, \pm 2 \pi / 3$, and $\pi$, at which the SAW propagation
direction merges to each mirror plane.
The amplitude takes the maximum at $|\sin 6 \varphi| = 1$, as shown for $\varphi = - \pi / 12$ in
Fig.~\ref{fig:5}, where we show how the $\phi'$ dependence of $W_{xy,0}$ for $\eta = 0$ is
changed by increasing $\lambda$.
The polar plot of the $\phi'$ dependence is symmetric in the spherical case $\lambda = 0$, which
becomes twisted and asymmetric for $\lambda \ne 0$.
Indeed, the indication of this behavior is inferred from the SAR measurements in SiC.
\cite{Hernandez-Minguez20}
The related comments are given in the next subsection.
We emphasize that the twisted behavior cannot be understood by the spherical approximation
because of the lack of information on the crystal structure; thus,
the consideration of the $C_{3v}$ symmetry is required for the V$_{\rm Si}$ spin--strain
interaction, as proposed by the present theory.
\par

\subsection{Asymmetric field-direction dependence of SAR}
Besides the single quantum spin transitions with $\Delta m_S = \pm 1$, the double quantum spin
transitions with $\Delta m_S = \pm2$ were also probed by the SAR measurements under a lower
magnetic field.
In Appendix~C, we describe the calculation of the latter transition rate and the effect of SAW
propagating in general directions on its asymmetric field-direction dependence.
Comparing the observed $\Delta m_S = \pm 2$ SAR data
\cite{Hernandez-Minguez20} with the calculated ones,
we find a possibility that the SAW propagation direction is tilted counterclockwise ($\varphi > 0$
measured from the $x$-axis).
To explain the asymmetric behavior of the $\Delta m_S = \pm 1$ SAR data,
\cite{Hernandez-Minguez20}
the coupling-strength ratio of the spin--strain interaction is required to satisfy $| h_{b'} / h_c | \ll 1$.
With these conditions, both the $\Delta m_S = \pm 1$ and $\Delta m_S = \pm 2$ data can be
simultaneously reproduced.
Thus, it is worth checking whether the SAW propagation direction is actually tilted from the
crystallographic mirror plane by more precise SAR measurements of the asymmetric
field-direction dependence.
\par

%%%%%%%%%%%%%%%%%%%%%%%%%%%%%%%%%%%%%%%%%%%%%%%%%%%%%%%%%%%%%%%%%%%%%%%%%%%%%%%%%%%%%%%%%%%%%%%%%%%
\section{Conclusion}
%%%%%%%%%%%%%%%%%%%%%%%%%%%%%%%%%%%%%%%%%%%%%%%%%%%%%%%%%%%%%%%%%%%%%%%%%%%%%%%%%%%%%%%%%%%%%%%%%%%
In this paper, we elaborated on the $S = 3/2$ V$_{\rm Si}$ spin--strain interaction considering the
$C_{3v}$ site symmetry for more detailed investigations of SAR under the rotation of a magnetic
field.
In our formulation, the ratios of coupling parameters, such as $h_{b'} / h_b$ and $h_{c'} / h_c$
in Eq.~(\ref{eqn:Ak}), can be evaluated from the field-direction dependence of the
$\Delta m_S = \pm 1$ transition rate, although the effects of various local strain fields are included
in the model Hamiltonian.
Importantly, the anisotropic SAR exhibits a marked maximum of the transition rate reflecting the
$C_{3v}$ site symmetry of V$_{\rm Si}$.
This provides more detailed information on the quadrupole components of the spin--strain
coupling beyond the spherical approximation.
In this study, we demonstrated how to determine the model parameters listed in Table~I
to evaluate the coupling-strength ratios for $C_{3v}$ from the observable SAR transition rates.
Considering the $\Delta m_S = \pm 2$ transition rate as well, we also pointed out the deviation of
the SAW propagation direction from the crystallographic mirror plane, which induces an additional
anisotropy in the SAR transition rates peculiar to the $C_{3v}$ symmetry.
These results will promote a thorough analysis of the spin--strain coupling to realize the acoustic
control of spin.
\par

It is also important to confirm the applicability of the coupling parameters determined by our
method to other experiments such as optical measurements using static stress fields.
\cite{Udvarhelyi18a}
When a uniaxial or shear stress is applied, the induced local strain field changes
the $S = 3/2$ energy levels of $V_{\rm Si}$.
Considering the spin--strain interaction Hamiltonian for the static strain field as presented here, we
can estimate the energy level shifts, which can be measured by electron spin resonance using
a microwave under a sufficiently strong stress to resolve each energy shift.
\par

We would like to make a comment on the coupling parameters $h_{b'}$ and $h_{c'}$ related to the
$zx$ and $yz$ quadrupole components.
As argued in Sect.~3.5, for the NV centers in diamond, it has been presumably considered that
the spin--strain coupling with $h_{b'}$ and $h_{c'}$ is less dominant than the coupling with
$h_b$ and $h_c$ related to the $x^2 - y^2$ and $xy$ quadrupole components.
It is still unclarified whether the former coupling parameters could be measurable.
On the other hand, the SAR measurement methods have been much more advanced for
V$_{\rm Si}$ than for the NV centers.
By applying our method to the SAR measurements, one can feasibly quantify the unknown
$h_{b'} / h_b$ and $h_{c'} / h_c$ parameters for $V_{\rm Si}$.
As mentioned in Sect.~3.7, the anisotropic SAR shows some important indications owing to
the effect of the $C_{3v}$ symmetry, and the dominant coupling parameters can be identified
by our theory when sufficient SAR data are accumulated.
\par

The Si vacancy centers have attracted more attention in relation to multiquantum transitions such
as multiphoton absorption processes in the $S = 3/2$ energy levels, which have recently been
observed by measurements of optically detected magnetic resonance.
\cite{Singh22}
This finding will stimulate a challenging investigation of multiphonon-driven quantum spin
transitions as the counterparts of multiphoton-driven transitions.
It is also intriguing to extend our theory to the SAR associated with multiphonon absorption transitions.
\cite{Koga20,Koga22}

%%%%%%%%%%%%%%%%%%%%%%%%%%%%%%%%%%%%%%%%%%%%%%%%
%%%%%%%%%%%%%%%%%%%%%%%%%%%%%%%%%%%%%%%%%%%%%%%%%%%
%\acknowledgment
%%%%%%%%%%%%%%%%%%%%%%%%%%%%%%%%%%%%%%%%%%%%%%%%%%%%%%%%%%%%%%%%%%%%%%%%%%%%%%%%%%%%%%%%%%%%%%%%%%%
%This work was supported by JSPS KAKENHI Grant Number 21K03466.
%\bigskip
%{\footnotesize
{\bf Acknowledgment}~~This work was supported by JSPS KAKENHI Grant Number 21K03466.
%}

\appendix
%%%%%%%%%%%%%%%%%%%%%%%%%%%%%%%%%%%%%%%%%%%%%%%%%%%%%%%%%%%%%%%%%%%%%%%%%%%%%%%%%%%%%%%%%%%%%%%%%%%
\section{Spin and quadrupole operators for spin-$3/2$}
On the $ | m_S \rangle$ basis for spin-3/2,
$\{ | 3/2 \rangle, |1/2 \rangle, | -1/2 \rangle , | -3/2 \rangle \}$,
the spin operators are explicitly given by the $4 \times 4$ matrices:
\begin{align}
& S_x = \frac{1}{2} \left(
\begin{array}{cccc}
0 & \sqrt{3} & 0 & 0 \\
\sqrt{3} & 0 & 2 & 0 \\
0 & 2 & 0 & \sqrt{3} \\
0 & 0 & \sqrt{3} & 0
\end{array}
\right),
\nonumber \\
& S_y = \frac{1}{2} \left(
\begin{array}{cccc}
0 & -i \sqrt{3} & 0 & 0 \\
i \sqrt{3} & 0 & -i 2 & 0 \\
0 & i 2 & 0 & -i \sqrt{3} \\
0 & 0 & i \sqrt{3} & 0
\end{array}
\right),
\nonumber \\
& S_z = \frac{1}{2} \left(
\begin{array}{cccc}
3 & 0 & 0 & 0 \\
0 & 1 & 0 & 0 \\
0 & 0 & -1 & 0 \\
0 & 0 & 0 & -3
\end{array}
\right).
\label{eqn:Sk}
\end{align}
Using these matrices, we drive the quadrupole operators as
\begin{align}
& O_u = \sqrt{3} \left(
\begin{array}{cccc}
1 & 0 & 0 & 0 \\
0 & -1 & 0 & 0 \\
0 & 0 & -1 & 0 \\
0 & 0 & 0 & 1
\end{array}
\right),
\nonumber \\
& O_v = \sqrt{3} \left(
\begin{array}{cccc}
0 & 0 & 1 & 0 \\
0 & 0 & 0 & 1 \\
1 & 0 & 0 & 0 \\
0 & 1 & 0 & 0
\end{array}
\right),
\nonumber \\
& O_{zx} = \sqrt{3} \left(
\begin{array}{cccc}
0 & 1 & 0 & 0 \\
1 & 0 & 0 & 0 \\
0 & 0 & 0 & -1 \\
0 & 0 & -1 & 0
\end{array}
\right),
\nonumber \\
& O_{xy} = \sqrt{3} \left(
\begin{array}{cccc}
0 & 0 & -i & 0 \\
0 & 0 & 0 & -i \\
i & 0 & 0 & 0 \\
0 & i & 0 & 0
\end{array}
\right),
\nonumber \\
& O_{yz} = \sqrt{3} \left(
\begin{array}{cccc}
0 & -i & 0 & 0 \\
i & 0 & 0 & 0 \\
0 & 0 & 0 & i \\
0 & 0 & -i & 0
\end{array}
\right).
\label{eqn:Qk}
\end{align}
Let us consider the unitary transformation from $\bS = ( S_x, S_y, S_z )$ to
$\bS_B = ( S_X, S_Y, S_Z)$,
\begin{align}
U^\dagger ( \be_\lambda \cdot \bS ) U = S_\lambda~~( \lambda = X, Y, Z ),
\end{align}
where the matrix representations of $S_X$, $S_Y$, and $S_Z$ in the $(XYZ)$ frame are given as
$S_x$, $S_y$, and $S_z$ in Eq.~(\ref{eqn:Sk}), respectively.
We choose
$\be_X = ( - \cos \theta \cos \phi, - \cos \theta \sin \phi, \sin \theta )$,
$\be_Y = ( \sin \phi, - \cos \phi, 0 )$, and
$\be_Z = ( \sin \theta \cos \phi, \sin \theta \sin \phi, \cos \theta )$
for the magnetic field $\bB \parallel Z$.
Accordingly, the explicit form of the unitary matrix $U$ is given as
\begin{align}
U = \left(
\begin{array}{cccc}
\chi^{-3} c c_{a+} & \chi^{-3} c \tilde{s} & \chi^{-3} s \tilde{s} & \chi^{-3} s c_{a-} \\
\chi^{-1} c \tilde{s} & \chi^{-1}  c c_{b-} & - \chi^{-1}  s c_{b+} & - \chi^{-1} s \tilde{s} \\
\chi s \tilde{s} & - \chi s c_{b+} & - \chi c c_{b-} & \chi c \tilde{s} \\
\chi^3 s c_{a-} & - \chi^3 s \tilde{s} & \chi^3 c \tilde{s} & - \chi^3 c c_{a+}
\end{array}
\right).
\label{eqn:US}
\end{align}
Here, each matrix element consists of the following terms:
\begin{align}
& \chi = e^{i \phi / 2},~~c = \cos \frac{ \theta }{2},~~s = \sin \frac{ \theta }{2},~~
\tilde{s} = \frac{ \sqrt{3} }{2} \sin \theta,
\nonumber \\
& c_{a \pm} = \frac{ 1 \pm \cos \theta }{2},~~c_{b \pm} = \frac{ 1 \pm 3 \cos \theta }{2}.
\end{align}
In the acoustically driven spin transitions, $S = 3/2$ states are coupled via the
quadrupole components $\{ O_u, O_v, O_{zx}, O_{xy},O_{yz} \}$.
For the rotation of a magnetic field, it is convenient to use the quadrupole operators in the
$(XYZ)$ frame, and the above unitary transformation gives
\begin{align}
& U^\dagger O_u U = - \frac{ 1 - 3 \cos^2 \theta }{2} O_U + \frac{ \sqrt{3} }{2} \sin^2 \theta O_V
\nonumber \\
&~~~~~~~~~~~~~~
+ \sqrt{3} \sin \theta \cos \theta O_{ZX},
\nonumber \\
& U^\dagger O_v U = \left( \frac{ \sqrt{3} }{2} \sin^2 \theta O_U
+ \frac{ 1 + \cos^2 \theta }{2} O_V - \sin \theta \cos \theta O_{ZX} \right)
\nonumber \\
&~~~~~~~~~~~~~~~~~~~~
\times  \cos 2 \phi
\nonumber \\
&~~~~~~~~~~~~~~~
+ ( - \cos \theta O_{XY} + \sin \theta O_{YZ} )  \sin 2 \phi,
\nonumber \\
& U^\dagger O_{zx} U = ( \sqrt{3} \sin \theta \cos \theta O_U - \sin \theta \cos \theta O_V
- \cos 2 \theta O_{ZX} )
\nonumber \\
&~~~~~~~~~~~~~~~~~~~~
\times \cos \phi
\nonumber \\
&~~~~~~~~~~~~~~
+ ( \sin \theta O_{XY} + \cos \theta O_{YZ} ) \sin \phi,
\nonumber \\
& U^\dagger O_{xy} U = \left( \frac{ \sqrt{3} }{2} \sin^2 \theta O_U
+ \frac{ 1 + \cos^2 \theta }{2} O_V - \sin \theta \cos \theta O_{ZX} \right)
\nonumber \\
&~~~~~~~~~~~~~~~~~~~~
\times \sin 2 \phi
\nonumber \\
&~~~~~~~~~~~~~~
+ ( \cos \theta O_{XY} - \sin \theta O_{YZ} ) \cos 2 \phi,
\nonumber \\
& U^\dagger O_{yz} U = ( \sqrt{3} \sin \theta \cos \theta O_U - \sin \theta \cos \theta O_V
- \cos 2 \theta O_{ZX} )
\nonumber \\
&~~~~~~~~~~~~~~~~~~~~
\times \sin \phi
\nonumber \\
&~~~~~~~~~~~~~~
+ ( - \sin \theta O_{XY} - \cos \theta O_{YZ} ) \cos \phi.
\label{eqn:OkOK}
\end{align}
Here, the matrices of the quadrupole operators $O_K$ ($K = U, V, ZX, XY, YZ$) are constructed by
the components of $\bS_B$ and given by the same representations of $O_k$
($k = u, v, zx, xy, yz$) in Eq.~(\ref{eqn:Qk}), respectively.
The unitary transformation $U^\dagger O_k U$ is represented by a linear combination of the five
components $O_K$. 
Accordingly, we obtain
$H_{\varepsilon, B} = U^\dagger H_{\varepsilon} U \equiv \sum_K A_{K, \varepsilon} O_K$
in Eq.~(\ref{eqn:HepB}), where $A_{K, \varepsilon}$ depends on the field direction represented by
$\theta$ and $\phi$.
\par

%%%%%%%%%%%%%%%%%%%%%%%%%%%%%%%%%%%%%%%%%%%%%%%%%%%%%%%%%%%%%%%%%%%%%%%%%%%%%%%%%%%%%%%%%%%%%%%%%%%
\section{Spherically symmetric spin--strain interaction model}
A spherically symmetric form of the spin--strain interaction is simply described by the following
Hamiltonian:
\begin{align}
& H_{\varepsilon}^{\rm s} = \xi \sum_{\alpha, \beta = x, y, z}
\varepsilon_{\alpha \beta} S_\alpha S_\beta
\nonumber \\
&~~~~~~
= \xi \left[ \frac{1}{2} ( \varepsilon_u O_u + \varepsilon_v O_v )
+ \varepsilon_{yz} O_{yz} + \varepsilon_{zx} O_{zx} + \varepsilon_{xy} O_{xy} \right],
\label{eqn:Hep-s}
\end{align}
where the coupling constant is given by a single parameter $\xi$, and the term of a bulk strain
component is not written.
In a similar form in Eq.~(\ref{eqn:Hep}), the strain-dependent coupling coefficients in
$H_{\varepsilon}^{\rm s} = \sum_k A_{k, \varepsilon}^{\rm s} O_k$ are written as
\begin{align}
& A_{u, \varepsilon}^{\rm s} = \frac{ \xi }{2} \varepsilon_u,~~
A_{v, \varepsilon}^{\rm s} = \frac{ \xi }{2} \varepsilon_v,~~
A_{zx, \varepsilon}^{\rm s} = \xi \varepsilon_{zx},
\nonumber \\
& A_{xy, \varepsilon}^{\rm s} = \xi \varepsilon_{xy},~~
A_{yz, \varepsilon}^{\rm s} = \xi \varepsilon_{yz}.
\label{eqn:Ak-s}
\end{align}
Comparing these $A_{k, \varepsilon}^{\rm s}$ and $A_{k, \varepsilon}$ in Eq.~(\ref{eqn:Ak}),
we can characterize the spherical symmetry of the spin--strain interaction by the coupling
parameters for $C_{3v}$ as
\begin{align}
h_a = h_b = h_{c'} = 2 \xi,~~h_c = h_{b'} =0.
\label{eqn:hs}
\end{align}
Thus, a deviation from the spherical symmetry is expressed by finite $h_c$ and $h_{b'}$.
For the $\pm x$ propagating SAWs we considered here, $h_c$ in $A_{v, \varepsilon}$ is related to
the $x^2 - y^2$ quadrupole component coupled to the $\varepsilon_{zx}$ field, whereas $h_{b'}$ in
$A_{zx, \varepsilon}$ is related to the $zx$ quadrupole component coupled to the $\varepsilon_v$ ($= \varepsilon_{xx}$) field.
Assuming that $\varepsilon_{xx}$ is real and $\varepsilon_{zx} \equiv - i \varepsilon''_{zx}$ for
a plane Rayleigh SAW, we treat $A_{v, \varepsilon}$ and $A_{zx, \varepsilon}$ as complex
numbers for $C_{3v}$,
\begin{align}
& | A_v | e^{i \theta_v}
= \frac{1}{ 2} \left( \frac{1}{2} h_b \varepsilon_{xx} + i h_c \varepsilon''_{zx} \right),
\nonumber \\
& | A_{zx} | e^{i \theta_{zx}}
=  \frac{1}{2} \left( - \frac{1}{2} h_{b'} \varepsilon_{xx} - i h_{c'} \varepsilon''_{zx} \right),
\end{align}
which lead to Eq.~(\ref{eqn:hbc}).
One can find that $\theta_v = 0$ and $| \theta_{zx} | = \pi / 2$ for the spherical spin--strain coupling
in Eq.~(\ref{eqn:hs}).
Thus, it is also useful to represent a deviation from the spherical symmetry by $| \theta_v |$ and
$( \pi / 2 - | \theta_{zx} | )$ instead of $h_c$ and $h_{b'}$, respectively.
\par

%%%%%%%%%%%%%%%%%%%%%%
\begin{figure}
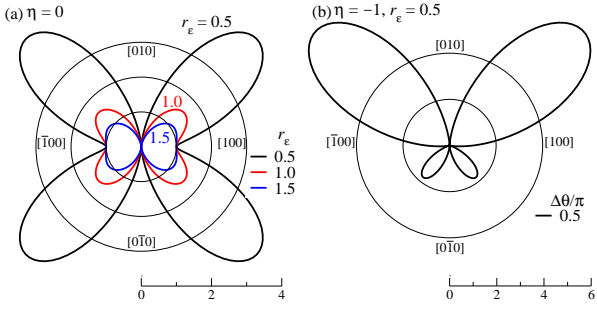

\begin{center}
\includegraphics[width=3.9cm,clip]{figb1a.eps}
\includegraphics[width=3.8cm,clip]{figb1b.eps}
\end{center}
\caption{
(Color online)
Field-direction $\phi$ dependence of the transition rate $W_{xy}$ in the spherical case.
(a) Polar representation of the normalized rate for $\eta = 0$.
The data are plotted for $r_{\varepsilon}  \equiv
2 \langle | \varepsilon''_{zx} \varepsilon_{xx} | \rangle / \langle \varepsilon_{xx}^2 \rangle = 0.5,
1.0$, and $1.5$.
(b) Polar representation of the normalized rate for $\eta = -1$, where $r_\varepsilon = 0.5$ and
$\Delta \theta / \pi \equiv ( \theta_{zx} - \theta_v ) / \pi = 0.5$.
}
\label{fig:b1}
\end{figure}
%%%%%%%%%%%%%%%%%%%%%%
In the spherical model, for the inplane ($xy$) field rotation, the $\phi$ dependence of the transition
rate $W_{xy}$ ($\eta = 0$) is almost the same as that observed in Fig.~\ref{fig:2}(b) for the
$C_{3v}$ site symmetry.
The only difference is that $\langle | a_{zx} | \rangle$ does not contain the coupling parameters but
is only strain-dependent as
\begin{align}
\langle | a_{zx} | \rangle
= \frac{ \langle | A_{zx, \varepsilon}^{\rm s} | | A_{v. \varepsilon}^{\rm s} | \rangle }
{ \langle | A_{v. \varepsilon}^{\rm s} |^2 \rangle }
= 2 \frac{ \langle | \varepsilon''_{zx} \varepsilon_{xx} | \rangle }{ \langle \varepsilon_{xx}^2 \rangle }
~~({\rm spherical}),
\end{align}
where Eq.~(\ref{eqn:Ak-s}) is used.
To compare the spherically symmetric and $C_{3v}$ symmetric models, we show $W_{xy}$ in
Fig.~\ref{fig:b1}(a) by replacing $\langle | a_{zx} | \rangle$ with
$r_\varepsilon \equiv 2 \langle | \varepsilon''_{zx} \varepsilon_{xx} | \rangle
/ \langle \varepsilon_{xx}^2 \rangle$ (see Fig.~\ref{fig:2}(b) for comparison).
Note that the V-shaped angular dependence cannot be obtained for
$r_\varepsilon \gtrsim 1$ in the spherical case ($h_b = h_{c'}$).
Applying a similar argument to $W_{xy}$ for $| \eta | = 1$ discussed in Sect.~3.3, we show a
polar plot for the spherical model in Fig.~\ref{fig:b1}(b).
This is the same as the plot for $\Delta \theta / \pi = 0.5$ in Fig.~\ref{fig:3}(b), where
$\langle | a_{zx} | \rangle = 0.5$ is replaced by $r_\varepsilon = 0.5$.

%%%%%%%%%%%%%%%%%%%%%%%%%%%%%%%%%%%%%%%%%%%%%%%%%%%%%%%%%%%%%%%%%%%%%%%%%%%%%%%%%%%%%%%%%%%%%%%%%%%
\section{Double quantum spin transition by SAW}
\subsection{Transition rate under inplane ($xy$) field rotation}
\begin{figure}
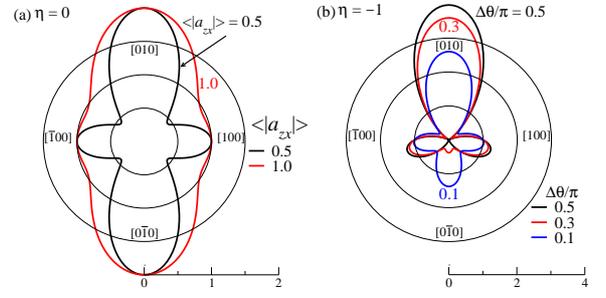

\begin{center}
\includegraphics[width=3.9cm,clip]{figc1a.eps}
\includegraphics[width=3.6cm,clip]{figc1b.eps}
\end{center}
\caption{
(Color online)
Polar representation of $W_{xy}^{(2)} / W_{xy, \phi = 0}^{(2)}$ for $\Delta m_S = \pm 2$ plotted
with respect to the field direction $\phi$.
(a) Standing SAW case ($\eta = 0$) plotted for $\langle | a_{zx} | \rangle = 0.5$ and $1.0$.
(b) Traveling SAW case ($\eta = -1$) plotted for
$\Delta \theta \equiv ( \theta_{zx} - \theta_v ) / \pi = 0.5, 0.3$, and $0.1$, where we choose
$\langle | a_{zx} | \rangle$ and fix $\langle a_u \rangle = 0$. 
}
\label{fig:c1}
\end{figure}

In our formulation, the $\Delta m_S = \pm2$ transition matrix element is calculated by the
quadrupole coupling operator $A_{V, \varepsilon} O_V + A_{XY, \varepsilon} O_{XY}$ in
Eq.~(\ref{eqn:HepB}).
By analogy with Eq.~(\ref{eqn:Wxy}), we represent the double quantum spin transition rate for the
inplane ($xy$) field rotation as
\begin{align}
& W_{xy}^{(2)} \propto \Big\langle \frac{3}{4}  | A_v |^2
\big\{ ( 3 a_u^2 + 2 \sqrt{3} a_u \cos \theta_v \cos 2 \phi + \cos^2 2 \phi )
\nonumber \\
&~~~~~~~~~~~~
+ 4 | a_{zx} |^2 \sin^2 \phi
\nonumber \\
&~~~~~~~~~~~~
+ 4 \eta | a_{zx} | [ \sqrt{3} a_u \sin \theta_{zx} + \sin ( \theta_{zx} - \theta_v ) \cos 2 \phi ] \sin \phi
\big\} \Big\rangle,
\label{eqn:w2xy}
\end{align}
where $a_u = A_u / | A_v |$ and $a_{zx} = | A_{zx} | / | A_v |$.
The parameters $\theta_v$ and $\theta_{zx}$ characterize the complex numbers as
$A_v = | A_v | e^{i \theta_v}$ and $A_{zx} = | A_{zx} | e^{i \theta_{zx}}$.
In Fig.~\ref{fig:c1}(a), we show the polar representation of the normalized transition rate
$W_{xy}^{(2)} / W_{xy, \phi = 0}^{(2)}$ for $\eta = 0$ to compare the data for $| a_{zx} | = 0.5$ and
$1.0$, which are the same parameters used in Fig.~\ref{fig:2}(b).
We also adjust the values of $\langle a_u \rangle$ and $\cos \theta_v$ to satisfy
$W_{xy, \phi = \pi / 2}^{(2)} / W_{xy, \phi = 0}^{(2)} = 2$ here.
As a function of $\phi$, these transition rates are explicitly written as
\begin{align}
& W_{xy, \eta = 0}^{(2)} ( \langle | a_{zx} | \rangle = 0.5 ) \propto \frac{3}{4}
( \cos^2 2 \phi + \sin^2 \phi ),
\label{eqn:w2xye0} \\
& W_{xy, \eta = 0}^{(2)} ( \langle | a_{zx} | \rangle = 1.0 ) \propto
\frac{3}{4} \left(1 + \frac{2}{3} \cos 2 \phi + \cos^2 2 \phi + 4 \sin^2 \phi \right).
\end{align}
Note that the four minima of $W_{xy}^{(2)}$ become prominent with the decrease in
$| a_{zx} |$ from unity, and this cross-shaped angular dependence indicates $| h_{c'} / h_b | < 1$
for the spin--strain coupling.
\par

%%%%%%%%%%%%%%%%%%%%%%
%%%%%%%%%%%%%%%%%%%%%%
In Eq.~(\ref{eqn:w2xy}), $W_{xy}^{(2)}$ for $\eta \ne 0$ shows an asymmetric $\phi$ dependence
with respect to the $x$-axis, which depends on $\Delta \theta$ ($\equiv \theta_{zx} - \theta_v$)
when we fix $\langle a_u \rangle \propto \langle A_u | A_v | \rangle = 0$.
In Fig.~\ref{fig:c1}(b), the normalized transition rate is plotted for $\eta = -1$ and
$\langle | a_{zx} | \rangle = 0.5$, where the parameters are changed in the same way as in
Fig.~\ref{fig:3}(b).
For $\Delta \theta = 0$, $W_{xy}^{(2)}$ is reduced to Eq.~(\ref{eqn:w2xye0}) and shows the
symmetric $\phi$ dependence for $\langle | a_{zx} | \rangle = 0.5$ in Fig.~\ref{fig:c1}(a).
In Fig.~\ref{fig:c1}(b), a prominent feature of $W_{xy, \eta = -1}^{(2)}$ is its
marked dependence on $\Delta \theta$ at $\phi = \pi /2$ (in the upper $xy$ plane).
\par

For the spherical model in Eq.~(\ref{eqn:Hep-s}), the $\phi$ dependence for $| \eta | = 1$
corresponds to $\Delta \theta / \pi = 0.5$.
For a large deviation $| \Delta \theta / \pi | \ll 0.5$ from the spherical symmetry, the $\phi$
dependence becomes symmetric, and this behavior is similar to that observed in Fig.~\ref{fig:2}(b)
for the single quantum spin transition.
\par

\subsection{SAW propagating in general directions}
In this subsection, we apply a similar argument in Sect.~3.6 to the field-direction
dependence of the $\Delta m_S = \pm 2$ transition rate.
First, we generalize Eq.~(\ref{eqn:Wc3vs}) for the $\Delta m_S = \pm 1$ transition rate as
\begin{align}
& W_{xy,0} \propto
2  + \lambda_c + \lambda_{b'} + ( 1 + \lambda_{b'} \cos 6 \varphi ) \cos 2 \phi'
\nonumber \\
&~~~~~~~~
- ( 1 + \lambda_c \cos 6 \varphi ) \cos 4 \phi'
\nonumber \\
&~~~~~~~~
+ \sin 6 \varphi ( \lambda_c \sin 4 \phi' - \lambda_{b'} \sin 2 \phi' ),
\label{eqn:Wc2vs2}
\end{align}
under the conditions of
$\langle | h_{b'} \varepsilon_{x' x'} |^2 \rangle
= \lambda_{b'} \langle | h_b \varepsilon_{x' x'} |^2 \rangle$,
$\langle | h_c \varepsilon_{x' z}'' |^2 \rangle
= \lambda_c \langle | h_{c'} \varepsilon_{x' z}'' |^2 \rangle$, and
$\langle | h_b \varepsilon_{x' x'} |^2 \rangle = \langle | 2 h_{c'} \varepsilon''_{x' z} |^2 \rangle$
in Eq.~(\ref{eqn:Wc3v}).
On the other hand, the $\Delta m_S = \pm 2$ transition rate for $\eta = 0$ is given by
\begin{align}
& W_{xy,0}^{(2)} \propto
\Big\langle \left[ h_a ( \varepsilon_{z z} - \varepsilon_{x' x'} / 2 )
+ ( h_b \varepsilon_{x' x'} / 2 ) \cos 2 \phi' \right]^2
\nonumber \\
&~~~~~~
+ 4 h_{c'}^2 \varepsilon_{x' z}''^2 \sin^2 \phi'
\nonumber \\
&~~~~~~
+ h_{b'}^2 \varepsilon_{x' x'}^2 \sin^2 (\phi' + 3 \varphi)
+ h_c^2 \varepsilon_{x' z}''^2 \cos^2 (2 \phi' + 3 \varphi) \Big\rangle.
\end{align}
Using the same conditions of the strain-dependent coupling parameters $\lambda_{b'}$ and
$\lambda_c$ in Eq.~(\ref{eqn:Wc2vs2}), we can reduce it to
\begin{align}
& W_{xy,0}^{(2)} \propto
\frac{5}{4} + \frac{5}{8} ( \lambda_c + \lambda_{b'} )
- \frac{1}{2}  ( 1 + 2 \lambda_{b'} \cos 6 \varphi ) \cos 2 \phi'
\nonumber \\
&~~~~~~~~
+ \frac{1}{4} ( 1 + \lambda_c \cos 6 \varphi ) \cos 4 \phi'
\nonumber \\
&~~~~~~~~
+ \frac{1}{4} \sin 6 \varphi ( 4 \lambda_{b'} \sin 2 \phi' - \lambda_c \sin 4 \phi' ).
\label{eqn:Wc2vs2m2}
\end{align}
Here, we have also assumed that
%$\langle | h_a ( \varepsilon_{z z} - \varepsilon_{x' x'} / 2 ) |^2 \rangle = 0$ and
%$\langle 4 h_a h_b \varepsilon_{x' x'} ( \varepsilon_{z z} - \varepsilon_{x' x'} / 2 ) \rangle
%= 1$
$\langle | h_a ( \varepsilon_{z z} - \varepsilon_{x' x'} / 2 ) |^2 \rangle = 0$ and
$\langle 4 h_a h_b \varepsilon_{x' x'} ( \varepsilon_{z z} - \varepsilon_{x' x'} / 2 ) \rangle
= \langle | h_b \varepsilon_{x' x'} |^2 \rangle$
to satisfy $W_{xy,0,\phi' = \pi / 2}^{(2)} / W_{xy,0,\phi' = 0}^{(2)} = 2$ for
$\lambda_c = \lambda_{b'} = 0$.
\par

%%%%%%%%%%%%%%%%%%%%%%
\begin{figure}
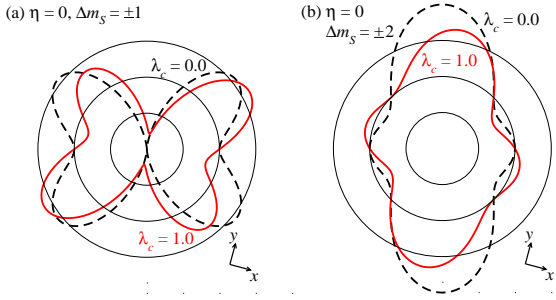

\begin{center}
\includegraphics[width=3.8cm,clip]{figc2a.eps}
\includegraphics[width=3.8cm,clip]{figc2b.eps}
\end{center}
\caption{
(Color online)
Polar representation of the normalized transition rate for $\eta = 0$.
(a) $W_{xy,0} / W_{xy,0,\phi' = 0}$ for $\Delta m_S = \pm 1$ in Eq.~(\ref{eqn:Wc2vs2}).
(b) $W_{xy,0}^{(2)} / W_{xy,0,\phi' = 0}^{(2)}$ for $\Delta m_S = \pm 2$ in
Eq.~(\ref{eqn:Wc2vs2m2}).
Here, the SAW propagation (horizontal) direction is tilted counterclockwise by $\varphi = \pi / 12$
from the $x$-axis of the crystal.
The angular $\phi'$ dependence is plotted for $\lambda_c = 0.0$ and $1.0$, where
$\lambda_{b'} = 0$ is fixed.
}
\label{fig:c2}
\end{figure}
%%%%%%%%%%%%%%%%%%%%%%
In Eqs.~(\ref{eqn:Wc2vs2}) and (\ref{eqn:Wc2vs2m2}), the odd function terms of $\phi'$ bring
about an asymmetric $\phi'$ dependence with respect to the $x'$- and $y'$-axes.
We find that $\sin 6 \varphi > 0$ reproduces the observed field-direction dependence of the
$\Delta m_S = \pm 2$ transition in SiC.
\cite{Hernandez-Minguez20}
Here, $\varphi > 0$ denotes that the SAW propagation direction is tilted counterclockwise.
On the other hand, $\sin 6 \varphi < 0$ is inferred from the $\Delta m_S = \pm 1$ data.
\cite{Hernandez-Minguez20}
Unless $\lambda_{b'}$ is extremely small in the $\sin 2 \phi'$ term, the observed field-direction
dependence can be identified in Fig.~\ref{fig:5}.
In other words, $\lambda_{b'} \ll \lambda_{c}$ is required to reproduce this asymmetric behavior
when $\varphi > 0$ is adopted for the SAW propagation direction.
\par

In Fig.~\ref{fig:c2}(a), we show the polar representation of $W_{xy,0}$ for $\Delta m_S = \pm 1$ in
Eq.~(\ref{eqn:Wc2vs2}) to demonstrate how the effect of the $\sin 4 \phi'$ term appears as the
asymmetric $\phi'$ dependence of $W_{xy,0}$.
Here, we compare the results of $\lambda_c = 0.0$ and $1.0$ in the absence of the $\sin 2 \phi'$
term ($\lambda_{b'} = 0$).
Figure~\ref{fig:c2}(b) shows the plot of $W_{xy,0}^{(2)}$ for $\Delta m_S = \pm 2$
in Eq.~(\ref{eqn:Wc2vs2m2}).
From these results, we can consider that the observed SAR data indicate $| h_{b'} / h_{c} | \ll 1$
for the spin--strain coupling, and a finite $h_c$ dominates the deviation from the spherical
symmetry in the spin--strain interaction.
\par

%%%%%%%%%%%%%%%%%%%%%%%%%%%%%%%%%%%%%%%%%%%%%%%
%%%%%%%%%%%%%%%%%%%%%%%%%%%%%%%%%%%%%%

%%%%%%%%%%%%%%%%%%%%%%%%%%%%%%%%%%%%%%%%%%%%%%%%%%%%%%%%%%%%%%%%%%%%%%%%%%%%%%%%%%%%%%%%%%%%%%%%%%%

\end{document}